\documentclass[iop,twocolappendix]{emulateapj}

\usepackage{amsmath,amsthm,amssymb}
\usepackage{etex,hyphenat}
\usepackage{natbib}

\usepackage{float}
\usepackage{graphicx}

\usepackage{xcolor}
\usepackage{hyperref}
\hypersetup{
   colorlinks,
   linkcolor={blue!88!black!80},
   citecolor={blue!88!black!80},
   urlcolor={blue!88!black!80}
}

\usepackage{bm}
\usepackage[T1]{fontenc}      
\newcommand{\ditto}[1][.4pt]{\textquotedbl}

\newcommand{\lya}{Ly$\alpha$}
\newcommand{\HeII}{He\,\textsc{ii}}

\newcommand{\CIII}{C\,\textsc{iii}]}

\newcommand{\LAGER}{\nohyphens{LAGER}}
\newcommand{\llya}{L$_{Ly\alpha}$}
\def\ergcm2s{\ifmmode {\rm\,erg\,cm^{-2}\,s^{-1}}\else
                ${\rm\,ergs\,cm^{-2}\,s^{-1}}$\fi}

\shorttitle{LAGER I: first results on reionization at $z\sim$ 7}

\shortauthors{Zheng, Zhen-Ya, et al.}

\begin{document}

\title{First Results from the Lyman Alpha Galaxies in the Epoch of Reionization (LAGER) Survey: Cosmological Reionization at $z\sim$ 7}
\author{
  Zhen-Ya Zheng\altaffilmark{1,2,3}, Junxian Wang\altaffilmark{4}, James Rhoads\altaffilmark{5,6}, 
  Leopoldo Infante\altaffilmark{2}, Sangeeta Malhotra\altaffilmark{5,6}, Weida Hu\altaffilmark{4}, 
  Alistair R. Walker\altaffilmark{7}, Linhua Jiang\altaffilmark{8}, Chunyan Jiang\altaffilmark{1,3,9}, 
  Pascale Hibon\altaffilmark{10}, Alicia Gonzalez\altaffilmark{5}, Xu Kong\altaffilmark{4}, 
  XianZhong Zheng\altaffilmark{11}, Gaspar Galaz\altaffilmark{2}, L. Felipe Barrientos\altaffilmark{2}.
}
\affil{
$^1$CAS Key Laboratory for Research in Galaxies and Cosmology, Shanghai Astronomical Observatory, 
Shanghai 200030, China; zhengzy@shao.ac.cn\\
$^2$Institute of Astrophysics and Center for Astroengineering, Pontificia Universidad Catolica de Chile, 
Santiago 7820436, Chile; linfante@astro.puc.cl \\
$^3$Chinese Academy of Sciences South America Center for Astronomy, Santiago 7591245, Chile \\
$^4$CAS Key Laboratory for Research in Galaxies and Cosmology, Department of Astronomy, 
University of Science and Technology of China, Hefei, Anhui 230026, China; jxw@ustc.edu.cn \\
$^5$School of Earth and Space Exploration, Arizona State University, Tempe, AZ 85287, USA; 
Sangeeta.Malhotra@asu.edu, James.Rhoads@asu.edu\\
$^6$Astrophysics Science Division, Goddard Space Flight Center, 8800 Greenbelt Road, Greenbelt, 
Maryland 20771, USA;\\
$^7$Cerro Tololo Inter-American Observatory, Casilla 603, La Serena, Chile\\
$^8$The Kavli Institute for Astronomy and Astrophysics, Peking University, Beijing, 100871, China\\
$^9$N\'ucleo de Astronom\'ia, Facultad de Ingenier\'ia y Ciencias , Universidad Diego Portales, Av. Ej\'ercito 441, Santiago, Chile\\
$^{10}$European Southern Observatory, Alonso de Cordova 3107, Casilla 19001, Santiago, Chile\\
$^{11}$Purple Mountain Observatory, Chinese Academy of Sciences, Nanjing 210008, China\\
}

\begin{abstract}
We present the first results from the ongoing \LAGER\ project (Lyman Alpha Galaxies in the Epoch of 
Reionization), which is the largest narrowband survey for $z\sim$ 7 galaxies to date.
Using a specially built narrowband filter NB964 for the superb large-area Dark-Energy Camera (DECam) 
on the NOAO/CTIO 4m Blanco telescope, \LAGER\ has  collected 34 hours NB964 narrowband imaging data in 
the 3 deg$^2$ COSMOS field.  We have identified 23 Lyman Alpha Emitter (LAE) candidates at $z=$ 6.9 in 
the central 2-deg$^2$ region, where DECam and public COSMOS multi-band images exist. The resulting luminosity 
function can be described as a Schechter function modified by a significant excess at the bright end (4 galaxies with 
L$_{Ly\alpha}\sim$ 10$^{43.4\pm0.2}$ erg\,s$^{-1}$). The number density at L$_{Ly\alpha}\sim$ 10$^{43.4\pm0.2}$ 
erg\,s$^{-1}$ is little changed from $z= 6.6$, while at fainter \llya\ it is substantially reduced. Overall, we see a 
fourfold reduction in \lya\ luminosity density from $z= 5.7$ to $6.9$. Combined with a more modest evolution of the 
continuum UV luminosity density, this suggests a factor of $\sim 3$ suppression of \lya\ by radiative transfer through 
the $z\sim 7$ intergalactic medium (IGM). It indicates an IGM neutral fraction $x_{HI}$ $\sim$ 0.4--0.6 
(assuming \lya\ velocity offsets of 100-200 km\,s$^{-1}$). The changing shape of the \lya\ luminosity function between $z\lesssim 6.6$ 
and $z=6.9$ supports the hypothesis of ionized bubbles in a patchy reionization at $z\sim$ 7. 
\end{abstract}


\section{Introduction}
\label{sec:intro}

In the last decade, much progress has been made in narrowing down the epoch of cosmological reionization to be 
between $z\sim 6$ \citep[\lya\ saturation in z $\sim$ 6 quasar,][]{Fan+2006} and $z \sim 9$ \citep[Polarization of 
CMB photons,][]{Planck2015XIII}. In between, quasars, gamma ray bursts (GRBs), and galaxies at $z\gtrsim$ 6 
are important to constrain the nature of reionization. However, the rarity of high-$z$ quasars \citep[with only one 
known at $z\gtrsim$ 7,][]{Bolton+2011}, and of high-$z$ GRBs with rapid followup, limit their application as 
reionization probes. In contrast, hundreds of galaxies at $z\gtrsim$ 6 have been reported from ground-based 
narrowband \lya\ line searches \citep[e.g.,][]{Ouchi+2010}, from space-based broadband Lyman break searches 
\citep[e.g.,][]{Bouwens+2015, Finkelstein+2015}, and from newly taken space-based grism spectroscopic surveys 
\citep{Tilvi+2016, Bagley+2017}. Their luminosities, number densities, clustering, and ionizing powers are essential 
to probe the epoch of reionization ({\it EoR}).

Because \lya\ photons are resonantly scattered by neutral hydrogen, Ly$\alpha$ emitters (LAEs) provide a sensitive, 
practical, and powerful tool to determine the epoch, duration, and inhomogeneities of reionization. With a sample of 
LAEs in {\it EoR}, the easiest \lya\ reionization test is the luminosity function (LF) test. \citet{RM01} first applied 
this test at $z$ = 5.7, then at 6.5 \citep{MR04, Stern+2005}. Subsequent surveys have confirmed and refined the 
$z$ = 6.5--6.6 neutral fraction measurement to $x_{HI}$ $\lesssim$ 0.3 \citep{Ouchi+2010, Kashikawa+2011},
and have found a significant neutral fraction increase from z = 6.6 to z = 7.3 \citep{Konno+2014}. 
Other implementations of \lya\ reionization tests include the \lya\ visibility test \citep{Jensen+2013, Schenker+2014, Faisst+2014}, 
i.e., the \lya\ fraction in LBGs in EoR, the volume test \citep[][Rhoads \& Malhotra 2017, in prep.]{MR06}
, i.e., each \lya\ galaxy is taken as evidence for a certain volume of ionized gas, and the clustering test, 
e.g., the signature of a patchy partially-ionized IGM is sought by looking for excess spatial correlations in 
the \lya\ galaxy distribution \citep{Furlanetto+2006, McQuinn+2007, Jensen+2013}.

Redshift $z\gtrsim 7$ is the frontier in \lya\ and re-ionization studies. Many searches for LAEs at z = 6.9--7.0
\citep{Iye+2006, Ota+2008, Ota+2010, Hibon+2011, Hibon+2012}, z = 7.2--7.3
\citep{Shibuya+2012, Konno+2014},  and z = 7.7 \citep{Hibon+2010, Tilvi+2010, Krug+2012} have yielded $\sim$
two dozen narrowband-selected candidates, but only 3 of these have been spectroscopically confirmed so
far \citep{Iye+2006, Rhoads+2012, Shibuya+2012}. In comparison, hundreds of LAEs have been found at lower redshift, 
at $z=6.6$ \citep[e.g., ][]{Hu+2010, Ouchi+2010, Kashikawa+2011, Matthee+2015}, 
and a large fraction of them have been spectroscopically confirmed \citep[i.e.,][]{Kashikawa+2011}. 
The comoving volumes of the searches for $z\gtrsim$ 7 LAEs are all less than 6$\times10^5$ cMpc$^3$, which 
are more likely to be affected by cosmic variance. In fact cosmic variance is more important for the observability 
of LAEs in patchy reionization. Further more, some of 
the surveys with 200\AA\ wide filters in the far-red inevitably \citep[e.g.,][]{Ota+2008,Ota+2010} include one or two weak OH emission lines, limiting their sensitivity.
Thus, there is an urgent need for a systematic survey at z $\sim 7$ with sufficient volume and depth to determine if the 
Ly$\alpha$ line is attenuated due to neutral IGM in a statistically significant way.

In this {\it Letter}, we report the discovery of a population of candidate $z\sim$ 7 \lya\ emitters in the first field (COSMOS) of 
the Lyman-Alpha Galaxies in the Epoch of Reionization (LAGER) survey. With 
LAGER-COSMOS,  we have the largest sample to date of candidate $z\sim$ 7 LAEs. 
In \S\ 2, we describe the observation and data reduction of LAGER, then introduce the candidates selection method.
In \S\ 3, we present the candidates and their \lya\ LF at $z\sim$ 7.
In \S\ 4, we discuss the implications of these discoveries for {\it EoR} at $z\sim$ 7.

\begin{figure*}[pt!]
\includegraphics[width=\linewidth]{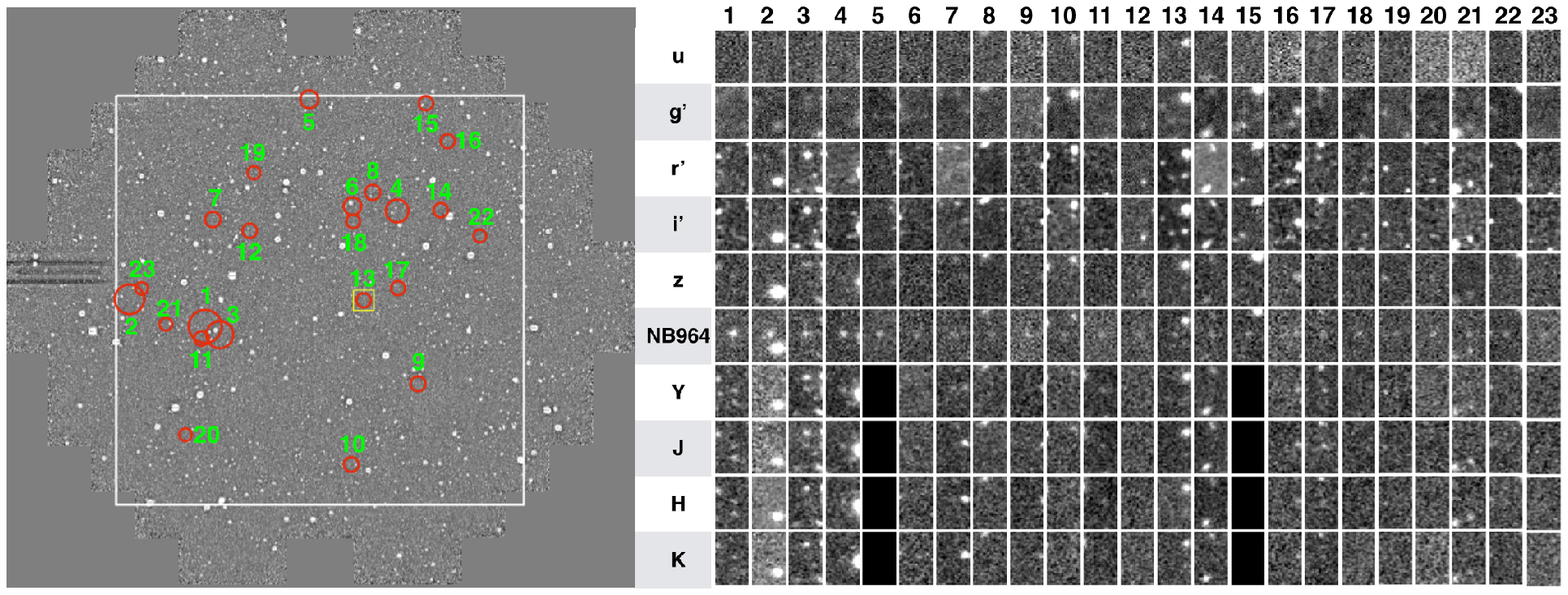} 
\caption{ \textbf{Left:}
The DECam NB964 COSMOS field showing uneven distribution of candidate $z\sim$ 7 LAEs. The DECam 
NB964 area is $\sim$3 deg$^2$, and the overlap region with Subaru Suprime-Cam multi-band images is the 
central 2 deg$^2$ area in the white box. We have selected 23 candidate $z\sim$ 7 LAEs, which are marked 
in red circles with aperture sizes scaled to their \lya\ line fluxes and normalized to $\sim$ 1 pMpc for the 
brightest LAE. The only spectroscopically confirmed LAE from previous narrowband survey \citep{Rhoads+2012} 
in this field is recovered and marked by a yellow box.  \textbf{Right:} Optical and NIR thumbnail 
images of the 23 candidate LAEs. We select to display images of DECam $u$,  Suprime-Cam $g'r'i'$, DECam $z$, 
DECam NB964, and UltraVISTA \citep[][DR3]{McCracken+2012} $YJHK_S$ bands. The stamp size is 10\arcsec$\times$5\arcsec.}
\label{fig:f-1}
\end{figure*}

\section{Survey Description and Data}
\label{sec:data}

\subsection{LAGER Survey}
The LAGER survey is the largest narrowband survey yet for LAEs at $z\gtrsim$ 7.  It is currently ongoing, using 
the Dark Energy Camera (DECam, with FOV $\sim$ 3 deg$^2$) on the NOAO-CTIO 4m Blanco telescope
together with an optimally designed custom
narrowband  filter NB964\footnote{Please see the filter information at NOAO website: 
\url{http://www.ctio.noao.edu/noao/content/Properties-N964-filter}} (Central wavelength $\sim$ 9642\AA, FWHM $\sim$ 90\AA).
LAGER is designed to select more than a hundred  z\,$\sim$\,7 LAEs over an area of 12 deg$^2$ in 4 fields ($8\times 10^6$ $\hbox{cMpc}^3$). 
Currently, LAGER has collected 47 hrs NB964 narrowband imaging in 3 fields (CDF-S, COSMOS, and DLS-f5)
taken during 9 nights in 2015 December, 2016 February, March and November.

The deepest NB964 imaging is done in COSMOS, where we have obtained 34 hours' NB964 exposure in a 3 deg$^2$ field. 
The exposure time per NB964 frame is 900s. Consecutive exposures were dithered by $\sim 100''$ so that chip gaps and 
bad pixels do not lead to blank areas in the final stacks. We also took $\sim$ 0.5--1 hrs z and Y bands exposure per field to 
exclude possible transients. There are deep archival DECam and Subaru broadband images in COSMOS.

\subsection{Data Reduction and Analysis}
\label{sec:photo}
We downloaded the reduced and calibrated DECam resampled images from the NOAO Science Archive, 
which were processed through the NOAO Community Pipeline \citep[version 3.7.0, ][]{Valdes+2014}. The 
individual frames were stacked following the weighting method in \citet{Annis+2014} with 
{\it SWarp} \citep[version 2.38.0,][]{Bertin2010}. 
The seeing of the final stacked LAGER-COSMOS narrowband image is $0.93''$.

The zero-magnitudes of the DECam images were calibrated to the Subaru Suprime-Cam broadband magnitudes 
of the stars in the public COSMOS/UltraVISTA $K_S$-selected catalog \citep{Muzzin+2013}. 
We estimated the image depth by measuring the root mean square (rms) of the background in blank places 
where detected ($>3\sigma$) signals were masked out. In a 2" diameter aperture, the 3$\sigma$ limiting 
AB magnitudes of the DECam images are [$u$, $g$, $r$, $i$, $z$, $Y$, NB964]$_{3\sigma}$\,=\,[26.2, 27.4, 26.9, 26.6, 26.1, 24.3, 25.6]. 
 Deep Subaru Suprime-Cam broad and narrow band images are available in the central 2-deg$^2$ area.
We downloaded the raw images from the archival server SMOKA \citep{Baba+2002}, and produced our own 
broad and narrow band stacked images \citep[follow-up work of][]{Jiang+2013}. Their 3$\sigma$ limiting 
magnitudes within a 2" diameter aperture are [$B$, $g'$, $V$, $r'$, $i'$, $z'$, NB711, NB816, NB921]$_{3\sigma}$
$\sim$ [27.9, 27.6, 27.2, 27.4, 27.3, 25.9, 26.0, 26.1, 26.2]. These Subaru Suprime-Cam images are deeper and 
have better seeing than the DECam images on average. 

We calculated the narrowband completeness via Monte Carlo simulations  with the {\it BALROG} software \citep{Suchyta+2016}.  The completeness fraction is defined 
as the {\it SExtractor} recovery percentage of the randomly distributed artificial sources in NB964 image as a 
function of narrowband magnitude. For point sources, the narrowband NB964 aperture magnitudes corresponding to the 80\%, 50\% 
and 30\% completeness fractions are 24.2, 25.0, and 25.3, respectively. For pseudo LAEs, we choose fake sources similar to that used in \citet{Konno+2017},
which have a S\'{e}rsic profile with the S\'{e}rsic index of $n$ = 1.5, and the half-light radius of $r_e$ $\sim$ 0.9 kpc (0.17 arcsec at $z=$ 6.9). The narrowband NB964 
aperture magnitudes corresponding to the 80\%, 50\% and 30\% completeness fractions are 24.3, 24.7, and 25.0 for pesudo LAEs, respectively.

\subsection{Selection Criteria for $z\sim$ 7 LAEs}
\label{subsec:selection}

Since Subaru broadband images are deeper and have better seeings than the DECam broadband images, we only 
apply selection criteria for z $\sim$ 7 LAEs in the central 2-deg$^2$ area covered by both Suprime-Cam and DECam. 
The selection criteria for $z\sim$ 7 LAEs include narrowband NB964 with $\geq$5-$\sigma$ detection 
(NB$964$ $\leq$ NB$964_{5\sigma} = 25.07$), narrowband NB964 excess over broadband z (z - NB964 $\geq$ 1.0)
\footnote{As NB964 lies to the edge of z band transmission curve, very red continuum sources may mimic emission line 
galaxies. We examine the z-NB vs z-Y plot for bright NB detected sources, and find red sources with z-Y $\sim$ 1.0 mag 
have  z-NB $\sim$ 0.4 mag, thus such effect is not important.}, 
and non-detections in both DECam $ugri$ and Suprime-Cam $Bg'Vr'i'$, N711, N816, N921 bands. We specially 
exclude broadband detections in the 1\arcsec.08 (4 pixel) diameter aperture to exclude marginal signals in these bands. 
156 sources passed the selection criteria, but a large fraction of them are fake sources (bleed trails, diffraction spikes, etc) 
after our visual check on NB964 image. With our team's tripartite visual check on both NB964 and broadband 
images, 27 candidate z $\sim$ 7 galaxies are selected. No transient was identified among them using available broadband images.

\section{Results}
\subsection{Candidate LAEs at $z\sim$ 7 in LAGER-COSMOS}
\label{subsec:z7laes}
We  estimate the \lya\ line fluxes and  EWs of the candidates using NB and  broadband photometry\footnote{UltraVISTA Y 
band  \citep[][DR3]{McCracken+2012} is used to directly estimate the UV continuum of LAGER LAEs. For those without UltraVISTA Y band 
coverage, we use DECam z band and follow the Appendix of \citet[][]{Zheng+2016} to estimate 
the \lya\ line fluxes and EWs. For broadband non-detections, $2\sigma$ upper limits are chosen. 
We assume a flat UV continuum of $f_v \propto v^{0}$ and the IGM absorption from \citet{Madau1995} in the 
above calculations. 
For candidates with $z$ or $Y$ band detection, such assumption holds within statistical uncertainty. 
}.
We further exclude 4 out of the 27 candidates with estimated line EW $<$ 10 \AA.

In total, 23 candidate z $\sim$ 7 LAEs, which have \lya\ line fluxes in the range of 
0.8--$7.8\times 10^{-17} \ergcm2s$ and rest-frame equivalent widths EW$_R$ $\geq$ 10\AA, are selected.
The corresponding observed \lya\ line luminosities are 5.4--$43.4\times 10^{42}$ erg\,s$^{-1}$. 
Their positions and their multi-band thumbnail images are plotted in Figure \ref{fig:f-1}.

The 4 brightest of them, with L(Ly$\alpha$) = 19.3--43.4 $\times$ 10$^{42}$ erg\,s$^{-1}$, are the most luminous 
candidate LAEs known at $z\gtrsim$ 7, thanks to the large survey volume of LAGER. We have confirmed 3 of the 
brightest LAEs at $z\sim$ 7 \citep{Hu+2017} with our recent Magellan/IMACS spectroscopic follow-up observations.

One of the candidates, J09:59:50.99+02:12:19.1, was previously selected and spectroscopically confirmed as 
a $z=6.944$  LAE with Magellan IMACS narrowband imaging and spectroscopy over a much smaller 
field \citep{Rhoads+2012}. It is well recovered with our selection using new DECam narrowband imaging.

\subsection{Ly$\alpha$ Luminosity Function at $z\sim$ 7}
\label{subsec:lyalfcalc}

We use the $V/V_\textrm{max}$ method to calculate the \lya\ LF at $z=$ 6.9. The formula below is adopted:
\begin{equation}
\Phi(L)d L = \sum_{L_i \in [L-\Delta L/2, L+\Delta L/2]} \frac{1}{V_{max} f_{comp}(NB_i)}.
\end{equation}
Here V$_{max}$ is simply a constant, the maximum survey volume for LAEs 
(V$_{max}$ = 1.26$\times$10$^6$ cMpc$^3$ calculated from sky coverage and NB964 central wavelength/FWHM),
and $f_{comp}(NB_i)$ is the detection completeness for sources with narrowband magnitude
m($\textrm{NB}_i$) interpolated in the narrowband detection completeness for fake LAEs  (see \S\ref{sec:photo}).

\begin{table}[htp]
\caption{\lya\ luminosity function at $z\sim$ 7 from LAGER-COSMOS field.}
\begin{tabular}{cccc}\hline\hline
log$_{10}$(L$_i$) & N $\pm$ $\Delta$ N$^{\ddag}$ & $\langle f_{\textrm comp.}\rangle$ & log$_{10}\Phi(L_i)$ \\
$_{(\Delta L=0.125)}$ & [L$_i\pm\frac{\Delta L}{2}]$ &    & [($d\textrm{log}_{10}L $)$^{-1}$ Mpc$^{-3}$] \\\hline
42.71 & 6$^{+3.6}_{-2.4}$ & 0.21 & -3.74$^{+0.20}_{-0.22}$  \\
42.84 & 7$^{+3.8}_{-2.6}$ & 0.44 & -4.00$^{+0.19}_{-0.20}$  \\
42.96 & 4$^{+3.2}_{-1.9}$ & 0.56 & -4.34$^{+0.25}_{-0.28}$ \\
43.09 & 2$^{+2.7}_{-1.3}$ & 0.80 & -4.80$^{+0.37}_{-0.45}$  \\
43.21 & $<2.6$ & 0.82 & $<$-5.10  \\
43.33 & 1$^{+2.3}_{-0.8}$ & 0.85 & -5.13$^{+0.52}_{-0.77}$ \\
43.46 & 2$^{+2.7}_{-1.3}$ & 0.88 & -4.84$^{+0.37}_{-0.45}$\\
43.59 & 1$^{+2.3}_{-0.8}$ & 0.90 & -5.15$^{+0.52}_{-0.77}$ \\\hline
\end{tabular}
\label{tab:lyalfz7}
\raggedright \\$^\ddag$: 1-$\sigma$ Poisson errors from \citet{Gehrels1986}.
\end{table}%

\begin{figure}[tbp]
\includegraphics[width=0.8\linewidth,angle=270]{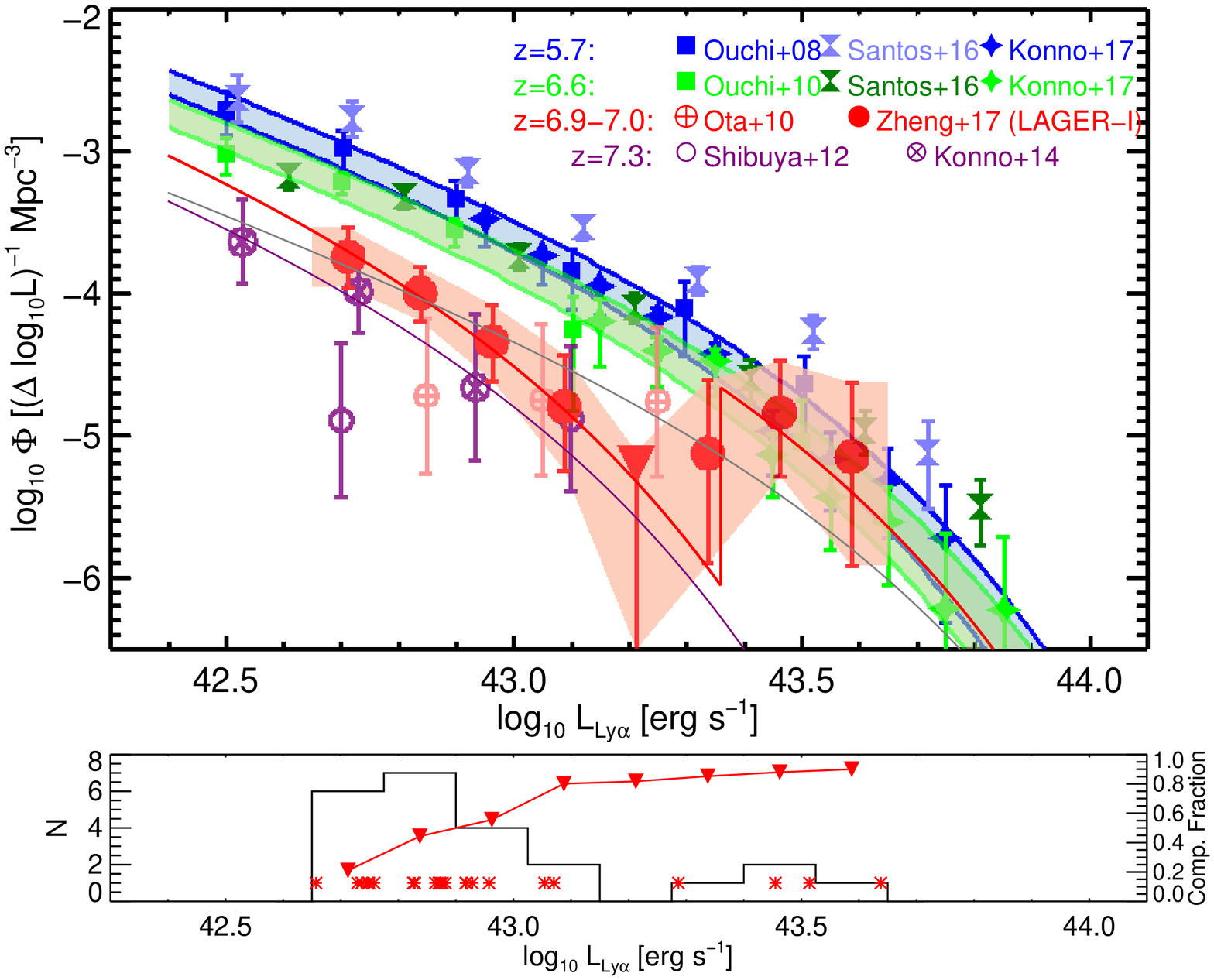} 
\caption{ {\it Top:} The \lya\ LFs of LAE surveys at $z\sim$ [5.7, 6.6, 7, 7.3]. 
The red filled circles show our results at $z\sim$ 7 from LAGER-COSMOS field.
References: \lya\ LFs at z = 5.7 from \citet{Ouchi+2008} (blue filled squares), \citet{Konno+2017} (blue filled stars), and
\citet{Santos+2016} (light-blue filled hourglass), at z = 6.6 from \citet{Ouchi+2010} (green filled squares), \citet{Konno+2017} 
(green filled stars), and \citet{Santos+2016} (dark-green filled hourglass), 
at $z\sim$ 7 from  this work (red filled circles) and \citet{Ota+2010} (pink empty circles with plus), and at z = 7.3 
from \citet{Shibuya+2012} (purple empty circles) and \citet{Konno+2014} (purple empty circles with cross). 
The colored regions in blue and green show the best Schechter fit with $\pm$1$\sigma$ error in $L*$ at $z\sim$ 5.7 and 6.6, 
respectively, from \citet{Konno+2017}. The solid and dashed lines in red (gray) show the best Schechter fits 
at $z\sim$ 6.9 with $L$ range 42.65 $\leq$ log$_{10}L$ $\leq$ 43.25 (42.65 $\leq$ log$_{10}L$ $\leq$ 43.65)
(see Table \ref{lyalfs}). 
A truncated Schechter function is also plotted (red solid line, see text for details) to measure the bright end LF bump.
The purple line shows the best Schechter fit at $z\sim$ 7.3 from \citet{Konno+2014}. 
 All these Schechter fits have a fixed faint-end slope $\alpha$ = -2.5. 
{\it Bottom:} The histogram of \lya\ luminosities of LAGER LAEs. The average completeness fraction 
for our LAEs interpolated from Section \ref{sec:photo}
 is also plotted as a red solid line with red triangles.
}      
\label{fig:f-3}
\end{figure}

The \lya\ LF of LAEs at $z\sim$ 7 in the LAGER-COSMOS field is listed in Table \ref{tab:lyalfz7}, and 
plotted in Figure \ref{fig:f-3} together with LAE LFs at $z\sim$ 5.7 -- 7.3 from literature.
The LF shows dramatic evolution not only in normalization, but also in the shape.

The relatively fainter end of our LF (42.65 $\leq$ log$_{10}L_{Ly\alpha}$ $\leq$ 43.25)  at $z\sim$ 7 can be fitted 
with a a single Schechter function in the form of 
  \begin{equation}
\Phi(L)d L = \Phi^*\left (\frac{L}{L^*}\right )^\alpha \exp\left (-\frac{L}{L^*}\right ) d \left(\frac{L}{L^*}\right)  ~~.
\label{eqn:schechter}
\end{equation}
Because our survey does not go far below $L^*$, we fix $\alpha=$ -2.5 in our fitting, which is 
suggested from recent studies on \lya\ LFs at $z=5.7$ and $z=6.6$ by \citet[][]{Konno+2017}. 
\citet[][]{Konno+2017} report the largest narrowband surveys to date for LAEs at  $z=5.7$ and $z=6.6$ taken with Subaru HSC
on the 14--21 deg$^2$ sky, and their best-fit Schechter function analyses also include the smaller but much deeper 
narrowband surveys taken with Subaru Suprime-Cam \citep[][]{Ouchi+2008,Ouchi+2010} at the corresponding redshifts.
Since the \lya\ LFs by \citet[][]{Konno+2017} are derived from the largest LAE samples at $z=5.7$ and $z=6.6$ with 
the widest luminosity range, and we have consistent selection and analysis methods,
we choose their \lya\ LFs for comparison in the following sections.
At $z>7$, we choose the \lya\ LF at $z=7.3$ by \citet{Konno+2014} for comparison.
The best-fit Schechter function parameters of these LFs are listed in Table \ref{lyalfs}, and plotted in the upper 
panel of Figure \ref{fig:f-4}. We find a large drop in $\Phi$ of \lya\ LF, compared with that at $z=$ 5.7 and $z=$ 6.6. 
In the relatively fainter end, our \lya\ LF at $z\sim$ 7 is in agreement within the 1$\sigma$ measurement uncertainties of that at $z=$ 7.3 .

More strikingly, we see a clear bump, i.e., significant excess to the Schechter function in the bright end 
of our $z\sim$ 7  LF (log$_{10}L_{Ly\alpha}$ $\gtrsim$ 43.25). It demonstrates that, while the space 
density of faint LAEs drops tremendously from $z=$ 5.7 and 6.6 to $z=$ 6.9, that of the luminous ones 
shows no significantly change.
Similarly but much less prominently,  the LF evolution between $z=$ 5.7 
and $z=$ 6.6 may also be differential,  with a significant decline at the faint end but not clear evolution 
at the bright end \citep{Matthee+2015,Santos+2016,Konno+2017}. 
In Fig. 2 we also plot a scaled-down and truncated version (red solid line at high luminosity end) of  \lya\ LF at $z=5.7$ from \citet{Konno+2017},
normalized by the UV luminosity density evolution between $z=5.7$ and $z=6.9$\footnote{Using $\rho^{UV}_{z=6.9}/\rho^{UV}_{z=5.7}$
interpolated from \citet{Finkelstein+2015}.}, which appears well consistent with our LF at the bright end within statistical uncertainties. 

\section{Discussion}

\subsection{The Evolution of Ly$\alpha$ LF in {\it EoR}}

\begin{figure}[tbp]
\includegraphics[width=0.76\linewidth,angle=270]{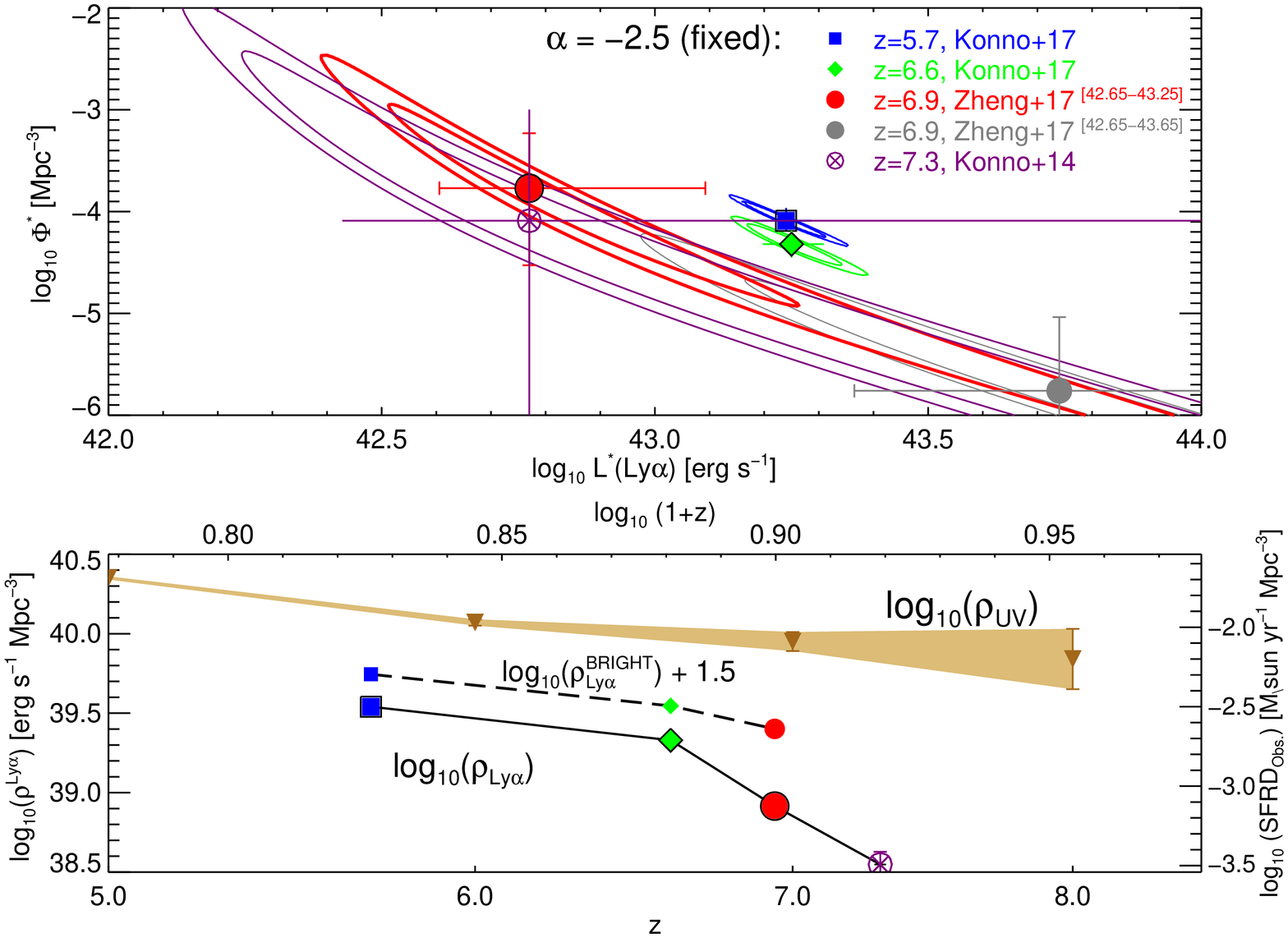} 
\caption{{\it Top:} The evolution of best-fit schechter parameters $L^*_{Ly\alpha}$ and $\Phi^*_{Ly\alpha}$ 
of \lya\ LFs at $z=$ [5.7, 6.6, 6.9, 7.3] with 1 and 2 $\sigma$ error contours.  
{\it Bottom:} Evolution of \lya\ and UV luminosity densities at $z\gtrsim$ 6. We plot the \lya\ luminosity densities 
integrated over the best-fit Schechter functions (42.4 $<$ log$_{10}L_{Ly\alpha}$ $<$ 44.0), and 
at the bright-end (43.3 $<$ log$_{10}L_{Ly\alpha}$ $<$ 43.7). Please see Table \ref{lyalfs} for the values. Note the bright luminosity density 
is scaled up by 1.5 dex. The yellow shaded region represents the UV luminosity densities for $L_{UV} >$ 0.03 $L^*_{UV}$ 
given by \citet{Finkelstein+2015} at $z\sim$ 5.0, 6.0, 7.0 and 8.0. The symbols are the same as those in Figure \ref{fig:f-3}.
}       
\label{fig:f-4}
\end{figure}

Cosmological reionization was well under way by $z\sim9$ \citep[][]{Planck2015XIII}, and 
ended by $z\sim 6$ \citep[][]{Fan+2006}. \lya\ galaxies at $z\gtrsim$ 6, e.g., from LAE surveys at 
$z$ = 5.7 \citep{Ouchi+2008,Santos+2016,Konno+2017}, at $z$ = 6.6 \citep{Ouchi+2010,Matthee+2015,Konno+2017}, at $z$ = 6.9 
\citep[our LAGER-COSMOS sample in this work and][]{Ota+2017}, and at $z$ = 7.3 \citep{Konno+2014}, 
are unique samples useful to probe both galaxy evolution and reionization.

\begin{table*}[htp]
\caption{Best-fit Schechter Parameters and \lya\ Luminosity Densities at $z\sim$ 5.7, 6.6, 6.9 and 7.3.} 
\begin{tabular}{cccc||cccc||c}\hline\hline
$z$ & Area & Volume &   $L$ Fitted Range &  log$_{10}$[$L^*_{Ly\alpha}$] & log$_{10}$[$\Phi^*$]  & log$_{10}$[$\rho_{Ly\alpha}$]$^\dag$ & log$_{10}$[$\rho_{Ly\alpha}^\textrm{Bright}$]$^\ddag$ & Reference \\ 
(1) & (2) & (3) & (4) & (5) & (6) & (7) &  (8)  & (9)\\\hline
      & deg$^2$ &   [cMpc$^3$] &   & [erg\,s$^{-1}$] & [Mpc$^{-3}$] & [erg\,s$^{-1}$\,Mpc$^{-3}$] & [erg\,s$^{-1}$\,Mpc$^{-3}$] & \\ \hline
       &   & &   & \multicolumn{4}{c}{ Two free parameters [$L^*_{Ly\alpha}$, $\Phi^*$] with fixed $\alpha$ = -2.5 } & \\  \hline
  5.7 &  13.8 & 1.16$\times$10$^7$ & 42.4--44.0 & 43.24$\pm$0.05 & -4.09$^{+0.13}_{-0.11}$ & 39.54$\pm$0.01   & 38.24    &{\scriptsize\citet{Ouchi+2008} + \citet{Konno+2017} }\\
  6.6 & 21.2 & 1.91$\times$10$^7$ &  42.4--44.0 &   43.25$^{+0.06}_{-0.05}$ & -4.32$\pm$0.13 & 39.33$\pm$0.02  & 38.05   &{\scriptsize\citet{Ouchi+2010} + \citet{Konno+2017} }\\
    6.9 & 2.0  &  1.26$\times$10$^6$ & \textbf{42.65--43.25}  &  42.77$^{+0.32}_{-0.16}$ & -3.77$^{+0.54}_{-0.76}$ &  $^{*}$38.92$^{+0.03}_{-0.04}$  & $^{*}$37.90   &   This study \\
     &      & &   \textbf{42.65--43.65}  &  43.74$^{+0.55}_{-0.38}$ & -5.76$^{+0.72}_{-0.81}$ &  38.77$\pm$0.05   & 37.82   &   This study \\ 
  7.3& 0.45   &  2.5$\times$10$^5$  & 42.4--43.0   & 42.77$^{+1.23}_{-0.34}$ & -4.09$^{+1.09}_{-1.91}$   & 38.55$^{+0.08}_{-0.07}$  & $^{**}$---  & \citet{Konno+2014} \\\hline\hline
\end{tabular}
\label{lyalfs}
\raggedright The best-fit parameters in Col. 5 and Col. 6 are derived by fitting the Schechter function with fixed $\alpha$ = -2.5 on the \lya\ LFs listed in Col. 9 with the $\chi^2$-statistics. The errors in Col. 5--7 are estimated from their 1-$\sigma$ confidence contours \citep[e.g.,][]{Coe2009} shown in the top panel of Fig. \ref{fig:f-4}. 
\\$^\dag$: \lya\ luminosity densities integrated over the best-fit Schechter functions at 42.4 $< \textrm{log}_{10}L_{Ly\alpha} <$ 44.0 (Col.-7). 
 Note the errors of \lya\ luminosity densities are likely underestimated because we do not consider the uncertainties of the faint-end slope in that calculation.
\\ $^\ddag$:  \lya\ luminosity densities integrated at the bright-end (43.3 $< \textrm{log}_{10}L_{Ly\alpha} <$ 43.7, Col.-8).\\
$^{*}$: The \lya\ luminosity densities in this row is calculated by integrating the truncated Schechter function plotted as the 
red solid curve in Figure \ref{fig:f-3} and discussed in Section \ref{subsec:lyalfcalc}. We use the \lya\ density in Col.-7 of this row to estimate the neutral IGM fraction at 
$z\sim$ 7 in Section \ref{xhi}. 
\\ $^{**}$: The survey for $z=7.3$ LAEs does not have large enough volume to cover the bright end. 
\end{table*}%

In this work, we detect for the first time a significant bump at the bright end of Ly$\alpha$ LF at $z\gtrsim$ 7.
Previous surveys for LAEs at $z\gtrsim$ 7 failed to reveal it as they covered much smaller volumes.

The existence of such bright end bump is consistent with the scenario proposed by \citet{HaimanCen2005}, 
that bright LAEs are less attenuated by a neutral IGM than faint LAEs, as the larger Str\"{o}mgren sphere 
surrounding luminous LAEs alleviates the neutral IGM absorption \citep[also see][]{Santos+2016,Konno+2017}. 
Therefore, the evolution in the bright end of the LF better reflects the intrinsic evolution of LAEs, while the faint 
end is controlled by the evolution of both galaxies and the IGM.

The bottom panel of Figure \ref{fig:f-4} shows the evolution of luminosity densities both for $Ly\alpha$ and 
UV photons. Compared to LBGs (yellow shaded region), an accelerated evolution of LAE $Ly\alpha$ densities 
from z = 6.6 (green diamonds) to z = 7.3 (purple cross), as reported by \citet{Konno+2014}, is clearly seen in 
Figure \ref{fig:f-4}. Since UV photons detected in LBGs are insensitive to the neutral hydrogen in EoR, 
\citet{Konno+2014} concluded that such rapid evolution in LAE LFs can be attributed to a large neutral IGM 
fraction at $z\sim$ 7.3. Our LAGER survey shows rapid evolution from $z=$ 6.6 to 6.9, and a smaller 
difference between $z=$ 6.9 and 7.3 (see Figure \ref{fig:f-4}).  It is possible that the evolution in the neutral 
fraction accelerated after $z=6.9$, though a smooth, monotonic neutral fraction evolution at $6.6\leq z\leq 7.3$ 
also fits the LAGER results and \citet{Konno+2014} measurements within their uncertainties. Note that the 
cosmic time from $z\sim$ 6.6 to 6.9 and from $z\sim$ 6.9 to 7.3 are approximately equal, which is about 50 Myr. 

Furthermore, compared to the evolution of \lya\ densities $\rho_{Ly\alpha}$ between redshifts of 5.7 and 6.9, the 
evolution in the bright end only $\rho_{Ly\alpha}^\textrm{Bright}$ follows a much smoother trend, somehow 
similar to that of the UV densities of LBGs (bottom panel of Figure \ref{fig:f-4}). This demonstrates 
that either the intrinsic LAE LF evolves moderately between z=5.7 and 6.9, or IGM attenuation plays a role even 
at the highest luminosity bin. Consequently, the dramatic and rapid evolution between $z=$ 6.6 and 6.9 in the faint 
end can be mainly attributed to the change of neutral IGM fraction. 
This evolution trend is in agreement with the \lya\ fraction test which shows the fraction of LBGs with visible \lya\ lines 
drops significantly at $z\gtrsim$ 7 \citep[e.g.,][]{Schenker+2014}.

\subsection{Neutral IGM fraction at $z\sim$ 7 with LAGER}
\label{xhi}

Following \citet{Ouchi+2010} and \citet{Konno+2014}, we compare the \lya\ luminosity densities at $z=$ 6.9 
(in EoR) and at $z=$ 5.7 (when reionization is completed) to estimate the effective IGM transmission factor 
$T^\textrm{IGM}_{Ly\alpha,z}$. Assuming the stellar population, interstellar medium (ISM), and dust are similar at $z$ = 6.9 and 
$z$ = 5.7,  we obtain 
\begin{equation}
T'_{z=6.9} = \frac{T^\textrm{IGM}_{Ly\alpha,z=6.9}}{T^\textrm{IGM}_{Ly\alpha,z=5.7}} = \frac{\rho^{Ly\alpha}_{z=6.9}/\rho^{Ly\alpha}_{z=5.7}}{\rho^\textrm{UV}_{z=6.9}/\rho^\textrm{UV}_{z=5.7}}.
\end{equation}
Here $\rho^\textrm{UV}$ and $\rho^{\textrm{Ly}\alpha}$ are UV and \lya\ luminosity densities, respectively. 
The ratio of UV luminosity densities $\rho^\textrm{UV}_{z=6.9}$/$\rho^\textrm{UV}_{z=5.7}$ = 0.63$\pm$0.09 is obtained 
from \citet{Finkelstein+2015}. We obtain the ratio of observed \lya\ luminosity densities 
(Col.-7 in Table \ref{lyalfs}) 
$\rho^{Ly\alpha}_{z=6.9}$/$\rho^{Ly\alpha}_{z=5.7}$ = 0.24$\pm$0.03.
Thus we estimate $T'_{z=6.9}$ = 0.38$\pm$0.07
\footnote{Noe the error of $T'_{z=6.9}$ is likely underestimated since we ignore the uncertainties from the faint-end slope of the \lya\ LFs.}
(c.f., $T'_{z=7.3}$ = 0.29 from \citealt{Konno+2014}  and $T'_{z=6.6}$ = 0.70$\pm$0.15 from \citealt{Konno+2017}).

Converting the \lya\ emission line transmission factor to neutral IGM fraction $x_{HI}$ is model dependent 
\citep[e.g.,][]{Santos2004, McQuinn+2007, Dijkstra+2007}. An important factor is the shift of the \lya\ line 
with respect to the systemic velocity, which is widely observed and may be explained by \lya\ radiative transfer 
in a galactic wind or outflow. The \lya\ velocity shift measurements of $z\sim$ 6--7 galaxies are very limited 
and are mostly around 100--200 km\,s$^{-1}$ (\citealt{Stark+2015,Stark+2017, Pentericci+2016},
but see two UV luminous galaxies with 400-500 km\,s$^{-1}$ reported by \citealt{Willott+2015}). 
With the analytic model of \citet{Santos2004}, assuming a \lya\ velocity shift of 0 and 360 km\,s$^{-1}$, 
the value of $T'_{z=6.9}$  = 0.38 corresponds to $x_{HI}$ $\sim$ 0.0 and 0.6, respectively.

By comparing our observed \lya\ LF at $z=6.9$ to that predicted with the radiative transfer simulations by 
\citet[][their Fig. 4]{McQuinn+2007}, we obtain $x_{HI}$ $\sim$ 0.40--0.60 at $z$ = 6.9. Similar results are 
obtained in the analytic studies considering ionization bubbles \citep{Furlanetto+2006, Dijkstra+2007}.  
We should note that these studies predicted a suppression of the luminosity function that is rather uniform 
across a wide range of luminosities \citep[e.g., Sec. 4 of][]{McQuinn+2007}. This suggests that the true 
distribution of ionized region sizes may differ appreciably from those used in literature \citep{Furlanetto+2006,Dijkstra+2007,McQuinn+2007}.
We conclude that the neutral hydrogen fraction is $x_\textrm{HI}$ $\sim$ 0.4--0.6  at $z=6.9$, 
where both the uncertainties in the IGM transmission factor calculation and the theoretical model predictions are considered. 
The LBG \lya\ fraction test by \citet{Schenker+2014} yields a similar neutral hydrogen fraction
($x_\textrm{HI} =$ 0.39$^{+0.08}_{-0.09}$) at $z\sim$ 7. 
Note smaller $x_\textrm{HI} =$ 0.3$\pm$0.2 at $z=$ 6.6 is obtained by \citet{Konno+2017} with Subaru HSC and Suprime-Cam surveys.

\subsection{Ionized Bubbles at $z\sim$ 7}

The bump at the bright end of the \lya\ LF is an indicator of large enough ($>$ 1 pMpc radius) ionized bubbles, where the Hubble 
flow 
can bring \lya\ photons out of resonance, thus leading to different evolution 
of \lya\ LF at the bright end and the faint end. 
The uneven distribution of these LAEs may indicate ionized bubbles in a patchy reionization at $z\sim$ 7.  
 Larger LAGER samples in future will more definitely 
establish whether the degree of clustering in Figure~\ref{fig:f-1} requires patchy reionization.

The origin of the ionized bubbles could be 
extrinsic or intrinsic, or both. 
From \citet[][]{MR06}, which gave the proper radius of a Str\"{o}mgren sphere 
and the Gunn-Peterson effect optical depth of a LAE with outflow velocity $\Delta V$ in {\it EoR}, we know 
that even the brightest LAE can not produce a large enough ionized bubble to effectively reduce the optical depth to $\tau\sim$ 1.
We thus would expect extrinsic contribution from additional satellite galaxies associated with the luminous LAEs. 
From Figure 1, we do see some such fainter nearby sources (e.g., the projected separations between LAE-1, 3 
and 11 are $\lesssim$ 3.4 arcmin, which corresponds to 1pMpc at $z=6.93$). 
Further deeper NB imaging would enable us to probe the bubbles by detecting more fainter galaxies associated with 
 the luminous LAEs.

The intrinsic explanation is that these luminous LAEs may be physically unusual objects, perhaps of a type not commonly found at lower redshift. 
For example, they could represent an AGN population that dominates over the ordinary star forming galaxies 
at lg$_{10}L_{Ly\alpha}$ $>$ 43.3 \citep[e.g.,][]{Matsuoka+2016}. Alternatively, they could be star forming galaxies where metallicities and/or dust
abundances are low enough to enable considerably larger \lya\ production or escape than is commonly seen at $z<$ 6 \citep[e.g.,][]{Stark+2015, Stark+2017, Mainali+2017}.
At $z\sim$ 7, \citet{Bowler+2014} find an excess of luminous LBGs where there should be an exponential cutoff at the bright-end of their UV LFs.
NIR and FIR spectroscopic follow-up of these luminous LBGs show very large \lya\ velocity offsets \citep[$>$300 km\,s$^{-1}$,][]{Willott+2015, Stark+2017}, which are significantly larger than the predicted velocity offsets for galaxies in EoR by \citet{Choudhury+2015}. Besides that, unusually strong carbon lines reported in two other LBGs at $z\sim$ 7 (CIV$\lambda$1548\AA\ in A1703-zd6 at $z$ = 7.045 and \CIII\ 1909\AA\ in EGS-zs8-1 at $z$ = 7.73) indicate unusual harder and more intense UV radiation than that of $z\sim$ 3 LBGs \citep{Stark+2015, Stark+2017}. In addition, a strong \HeII\ line has been reported in the brightest \lya\ emitter at $z$ = 6.6 (CR7), suggesting it could be powered by either Pop. III stars \citep{Sobral+2015} or perhaps accretion on to a direct collapse black hole \citep{Pallottini+2015}.

Future deeper NB964 imaging and IR (NIR and/or FIR) spectroscopic observation will help us to determine the 
 nature of these luminous LAEs, and probe the ionized bubbles in a patchy reionization at $z\sim$ 7.

\section{Conclusion}
In this letter, we report the first results of our LAGER project, the discovery of 23 (22 new) candidate LAEs 
at $z\sim$ 7. This is the largest sample to date of candidate LAEs at $z\gtrsim$ 7. Further more, thanks to 
the large survey volume of LAGER, we find 4 most luminous candidate LAEs at $z\sim$ 7 with L(\lya) = 
19--43$\times$10$^{42}$ erg\,s$^{-1}$. Compared to previous \lya\  LFs at $z\gtrsim$ 6, 
the \lya\ LF of LAGER LAEs at $z\sim$ 7 shows different evolution at the faint-end and 
at the bright-end, which indicates a large neutral hydrogen fraction $x_\textrm{HI}$ $\sim$ 0.4--0.6 and 
the existence of ionized bubbles at $z\sim$ 7. Our findings support the patchy reionization scenario at $z\sim$ 7.

\acknowledgments

We thank the anonymous referee for careful and helpful comments which improve the manuscript. 
We acknowledge financial support from National Science Foundation of China (grants No. 11233002 \& 11421303) 
and Chinese Top-notch Young Talents Program for covering the cost of the NB964 narrowband filter. J.X.W. thanks 
support from National Basic Research Program of China (973 program, grant No. 2015CB857005), and CAS Frontier 
Science Key Research Program QYCDJ-SSW-SLH006. Z.Y.Z acknowledges supports by the China-Chile Joint Research 
Fund (CCJRF No. 1503) and the CAS Pioneer Hundred Talents Program (C). L.I. is in part supported by CONICYT-Chile grants Basal-
CATA PFB-06/2007, 3140542 and Conicyt-PIA-ACT 1417. C.J. acknowledges support by Shanghai Municipal Natural 
Science Foundation (15ZR1446600)

We thank Materion company for the manufacture of the NB964 filter, which made the \LAGER\ project 
possible. We greatly appreciate the kind support from staffs at NOAO/CTIO to make our observations successful. 

Based on observations at Cerro Tololo Inter-American Observatory, National Optical Astronomy Observatory (NOAO 
PID: 016A-0386, PI: Malhotra, and CNTAC PIDs: 2015B-0603 and 2016A-0610, PI: Infante), which is operated by the 
Association of Universities for Research in Astronomy (AURA) under a cooperative agreement with the National Science Foundation. 
Based in part on data collected at the Subaru Telescope and obtained from the Subaru-Mitaka-Okayama-Kiso Archive System (SMOKA), 
which is operated by the Astronomy Data Center, National Astronomical Observatory of Japan.

This project used data obtained with the Dark Energy Camera (DECam), which was constructed by the Dark Energy Survey 
(DES) collaboration. Funding for the DES Projects has been provided by the DOE and NSF (USA), MISE (Spain), STFC (UK), 
HEFCE (UK). NCSA (UIUC), KICP (U. Chicago), CCAPP (Ohio State), MIFPA (Texas A\&M), CNPQ, FAPERJ, FINEP (Brazil), 
MINECO (Spain), DFG (Germany) and the collaborating institutions in the Dark Energy Survey, which are Argonne Lab, UC Santa 
Cruz, University of Cambridge, CIEMAT-Madrid, University of Chicago, University College London, DES-Brazil Consortium, 
University of Edinburgh, ETH Zurich, Fermilab, University of Illinois, ICE (IEEC-CSIC), IFAE Barcelona, Lawrence Berkeley 
Lab, LMU Munchen and the associated Excellence Cluster Universe, University of Michigan, NOAO, University of Nottingham, 
Ohio State University, University of Pennsylvania, University of Portsmouth, SLAC National Lab, Stanford University, University 
of Sussex, and Texas A\&M University.

{\it Facilities:} 
\facility{$Blanco$ (DECam)},
\facility{$Subaru$ (Suprime-Cam)}

\end{document}